\documentclass[journal=jacsat,manuscript=article]{achemso}

\usepackage[version=3]{mhchem} % Formula subscripts using \ce{}

\usepackage{xcolor}
\usepackage{float}
\usepackage{lscape}
\usepackage{cleveref}

\newcommand\ket[1]{|#1\rangle }
\newcommand\bra[1]{\langle #1|}
\newcommand\braket[2]{\langle {#1}|{#2}\rangle }
\SectionNumbersOn
\setkeys{acs}{articletitle=false}

\newcommand{\MicrosoftR}[0]{Microsoft Quantum, Redmond, Washington 98052, USA}
\newcommand{\MicrosoftZ}[0]{Microsoft Quantum, 8038 Z\"urich, Switzerland}
\author{Hongbin Liu}
\affiliation{\MicrosoftR}
\author{Guang Hao Low}
\affiliation{\MicrosoftR}
\author{Damian S. Steiger}
\affiliation{\MicrosoftZ}
\author{Thomas H\"aner}
\affiliation{\MicrosoftZ}
\author{Markus Reiher}
\affiliation{Laboratorium f\"ur Physikalische Chemie, ETH Z\"urich, Vladimir-Prelog-Weg 2, 8093 Zürich, Switzerland}
\email{ markus.reiher@phys.chem.ethz.ch }
\author{Matthias Troyer}
\affiliation{\MicrosoftR}
\email{mtroyer@microsoft.com}

\title{Prospects of Quantum Computing\\ for Molecular Sciences
}

\keywords{American Chemical Society, \LaTeX}

\begin{document}

\begin{abstract}
Molecular science is governed by the dynamics of electrons and atomic nuclei, and by their interactions with electromagnetic fields. A faithful physicochemical understanding of these processes is crucial for the design and synthesis of chemicals and materials of value for our society and economy. Although some problems in this field can be adequately addressed by classical mechanics, many demand an explicit quantum mechanical description. Such quantum problems require a representation of wave functions that grows exponentially with system size and therefore should naturally benefit from quantum computation on a number of logical qubits that scales only linearly with system size.
In this perspective, we elaborate on the potential benefits of quantum computing in the molecular sciences, i.e., in molecular physics, chemistry, biochemistry, and materials science.
\end{abstract}

\section{Introduction}

Quantum computing promises exponential speedups over traditional computing for certain  computational problems 
%[\todo{cite reviews}]
\cite{shor1999polynomial,abrams1999quantum,harrow2009quantum,childs2003exponential,kassal2008polynomial,montanaro2016quantum,cao2019quantum,emani2021quantum,outeiral2021prospects}. 
Although recent developments in quantum hardware and algorithms
\cite{low2019hamiltonian,vonberg2020,lee2020even,pino2020demonstration,google2020hartree,Google19,CAS20}
are impressive, a potential quantum advantage has only been demonstrated for toy problems. \cite{Google19,CAS20, IBM2019}.
Looking to the future, it will be crucial to demonstrate a quantum advantage for problems of scientific or industrial relevance that are legitimately intractable by traditional computing. 
This prospect justifies the enormous financial investment needed to realize universal quantum computation.

An exponential rather than polynomial quantum speedup is the natural target for current developments as it allows one to clearly define an application that hits a wall in traditional computing due to the curse of dimensionality. However,
it is important to note that exponential speedup is promised only by a very limited number of quantum algorithms \cite{zoo2021}, of which one is the simulation of quantum systems \cite{abrams1999quantum,nielson2000quantum}. Further quantum algorithms have been developed in recent years that achieve a polynomial speedup. However, as a single classical GPU chip can have a factor of $10^{10}$ better performance in bit and floating-point operations than a single quantum chip,\cite{haner2020assertion,troyer2021simons} the latter will be more difficult to show a clear advantage over traditional computing.

Molecular science is a key application area for quantum computing as the quantum dance of electrons and nuclei in molecules occurs on the nanometer scale, and must be described by a quantum model. However, despite this tiny scale, molecular events can have a dramatic macroscopic impact as highlighted by key chemical processes in nature and industry such as (i) nitrogen fixation and fertilizer production, (ii) photosynthetic light harvesting and photovoltaic cells, and (iii) bio-macromolecular chemistry and polymer materials.

While the possibility to describe molecular phenomena in terms of quantum algorithms was shown some time ago \cite{lloyd1996,aspuru-guzik2005,veis2010,cao2019quantum,bauer2020quantum}, we demonstrated in 2016 that quantum computing may actually have the potential to solve relevant chemical problems such as nitrogen fixation catalysis because the resources required are feasible in terms of the size of a machine and the time scale required for a calculation\cite{reiher2017}. Recently, we extended this work with respect to further algorithmic development and application range \cite{vonberg2020}. Whereas our work has been based on theoretical analyses only, actual quantum computations have already been carried out in pioneering work for molecular toy systems
\cite{o2019quantum,o2019calculating,nam2020ground,kandala2017hardware,google2020hartree,kawashima2021efficient}. At the same time, traditional algorithms in classical computing have become very mature and efficient in the past decades and present a clear challenge as competing approaches to quantum computing.

In this perspective, we provide a broader view on typical problems in the molecular sciences that are important targets for quantum algorithms and we also discuss traditional approaches to tackle them. We attempt to assess what is currently known about the potential of quantum algorithms to replace traditional approaches on both near-term and future quantum devices. We also elaborate on the criteria that eventually allow one to
assess the advantage of quantum computing over traditional computing in this context.

\section{Computational Challenges in Molecular Science}

We begin our discussion with an overview of key problems in the physical description of phenomena in chemistry and materials science. Obviously, we have to condense these broad fields to key physical effects connected to relevant applications. However, we emphasize that due to the general nature of physical models, our analysis can be generalized to similar problems (not only in the molecular sciences) in which the same type of physical modelling is applied.

\Cref{tab:overview} provides an overview on relevant problems
in the molecular sciences: 

(1) Molecular structure prediction comprises static as well as dynamic procedures that assign an energy to a given set of Cartesian nuclear or ionic coordinates and hence make them comparable in terms of this energy, which eventually allows one to search for the lowest-energy structure.

(2) Related to the energy assignment in (1) is the sampling of very many structures of a system under macroscopic constraints such as constant temperature, volume, and particle number in order to access microstate energies that are relevant for the partition function and hence for thermodynamic quantities such as the free energy.

(3) Also related to (1) is tracking the energy along a structural change that describes a chemical reaction, which requires a quantum description of the electrons in order to accurately adapt to any of the nuclear scaffolds that might be visited along such a reactive trajectory.

(4) Naturally, this may also occur in an electronically excited state accessible by light irradiation, which requires the calculation of more than the lowest energy eigenvalue of the electronic Schr\"odinger equation.

(5) Whereas the preceding problems typically rely on the stationary Schr\"odinger equation, some processes may require explicit dynamics of the elementary particles (electrons and nuclei in this case) and, therefore, their quantum dynamics must be studied explicitly.

(6) In the last column of \cref{tab:overview} we added a branch of computational science that is rather unrelated to a specific underlying mechanical theory: data-driven cheminformatics which has been propelled recently by developments in machine learning and artificial intelligence, for which traditional as well as quantum algorithms have been advanced.

Key to the understanding of all of these application areas is that the molecular processes are reduced to the dynamics of electrons and nuclei or to that of entities composed of them (i.e., atoms and molecules). The energy assignment is typically done in terms of the electronic energy emerging from the Born-Oppenheimer approximation that freezes out the motion of the nuclei, which are much heavier than electrons. This energy can either be supplemented with quantum corrections for the motion of the atomic nuclei by solving the Schr\"odinger equation including the nuclei or through Newtonian dynamics in a classical approximation. In the latter context, the electronic energy may be efficiently approximated for certain problems by a force-field (FF) to enhance computational efficiency and sampling -- in particular, for large, heterogeneous atomistic structures. 

In traditional computing, computational efficiency often compromises accuracy. However, depending on the question to be answered by computer simulation, a computational result may have modest accuracy requirements. An issue in this context is that these algorithms involve uncontrolled approximations and thus do not supply rigorous error bounds. We emphasize this aspect of traditional methods because rigorous error estimates are available in certain quantum algorithms.
In \cref{tab:overview}, we provide typical accuracy requirements for the target quantities. 

\begin{landscape}
\begin{table}[ht]
\footnotesize
\begin{tabular}{|p{33mm}|p{29mm}|p{29mm}|p{32mm}|p{34mm}|p{30mm}|p{29mm}|}
\hline
Applications & Molecular structure prediction \& exploration & Biochemical processes (e.g. drug-molecule protein docking) & Ground state chemistry (e.g. catalysis) & Photochemistry (e.g. photosynthesis) & Complex   dynamics (e.g. charge dynamics) & Cheminformatics \\ \hline
Chemical physics & Forces on atom & Thermodynamics & Kinetics & Spectroscopy & Electronic \& nuclear dynamics & Data-driven, physics-inspired, cost-function optimization \\ \hline
Physical quantity to be calculated & Energy gradient & Free energy (difference) & Reaction \& activation energies & Excitation energies & Autocorrelation functions & Universal applicability \\ \hline
Accuracy   (Hartree atomic units) & $10^{-3}\sim 10^{-4}$  & $10^{-3}\sim 10^{-4}$ & $10^{-3}\sim 10^{-4}$ & $10^{-3}$ & purpose dependent & purpose dependent \\ \hline
Mechanical theory & Effectively classical dynamics of nuclei/ions& Effectively classical dynamics of nuclei/ions & Electronic Schr\"odinger equation for ground states & Electronic Schr\"odinger equation for excited states & Time-dependent electronic and nuclear Schr\"odinger equations& Agnostic to the underlying mechanical theory \\ \hline
State-of-the-art traditional competitors & DFT, QM/MM & FF,DFT,QM/MM & CCSD(T), CASSCF, DMRG-CI/SCF, FCIQMC/SCF, MR-PT2, MRCI+Q & EOM-CCSD/CC3, DMRG-CI/SCF, MR-PT2, MRCI+Q,  & MCTDH & Neural Network \\ \hline
Routine traditional competitor & DFT, FF & FF & DFT, MP2 & TD-DFT, ADC(2) &  Surface hopping &  \\ \hline
Quantum algorithms & Quantum   search & Quantum  Metropolis-Hastings & QPE, VQE & QPE, VQE & Hamiltonian  simulation & Quantum   Machine Learning \\ \hline
Quantum   speedup & Quadratic  &  Polynomial & Exponential  & Exponential  & Exponential  & Unknown \\ \hline
\end{tabular}
\caption{Overview of potential application areas for quantum algorithms alongside with information about relevant physical target quantities, underlying theoretical foundations, and some of the traditional algorithms to compete with. An introduction and explanation of the acronyms is given in the text.}
\label{tab:overview}
\end{table}
\end{landscape}

\section{Principles of Quantum Many-Body Methods}
%Energy evaluation
The ultimate goal of computational molecular science is to solve the time-dependent Schr\"odinger equation,
\begin{equation} \label{eq:tdse}
    H\ket{\Psi(\vec{r}_1,...,\vec{r}_N, t)} = i\hbar\frac{\partial}{\partial t}\ket{\Psi(\vec{r}_1,...,\vec{r}_N, t)},
\end{equation}
accurately, where $\ket{\Psi}$ may be taken as the $N$-electron wave function after introducing the Born-Oppenheimer approximation. The time dependence can be treated separately as long as the Hamiltonian does not depend on time,
\begin{equation} \label{eq:tise}
    H\ket{\Psi(\vec{r}_1,...,\vec{r}_N)} = E\ket{\Psi(\vec{r}_1,...,\vec{r}_N)},
\end{equation}
and we are most interested in the electronic energy $E$ (and in many cases, in the ground state energy, $E_0$) of a molecular system. \Cref{eq:tise} is hard to solve exactly, as it is a $3N$-dimensional linear second-order partial differential equation (PDE) for $N$ electrons. A standard approach to solve such an equation is through basis set expansion. It is the dimension of this many-electron basis function space that scales exponentially with the size of the system, e.g., with the particle number $N$. 

In the past decades, tremendous achievements have been made in developing approximate traditional methods for solving this equation for chemical systems, aiming to balance accuracy with computational feasibility. Methods that scale polynomially, such as density functional theory (DFT) \cite{hohenberg1964inhomogeneous,kohn1965self,kohn1996density} and coupled-cluster (CC) \cite{purvis1982full,piecuch2002recent,bartlett2007coupled}, have been widely used to determine approximations to the ground state energy of chemical systems. 

While these methods are rooted in different foundations, some common ground has been established to make \cref{eq:tise} solvable for chemical systems: 1) We use a finite one-electron basis, \textit{e.g.}, atomic Gaussian basis functions, for the construction of the many-electron basis states and 2) a single Slater determinant (antisymmetrized Hartree product) is an example of such a many-electron basis state and usually taken as a starting point to systematically approximate the many-electron wave function.

Among approximate classical methods, DFT is used most prevalently for evaluating ground state energies of molecules involving any elements from the periodic table and homogeneous materials such as metals and semiconductors. Its relatively low scaling of around $\mathcal{O}(m^3)$ 
with respect to the number $m$ of one-electron basis functions enables routine calculations of chemical systems with up to about a thousand atoms. The calculated ground state energies can be directly used to answer questions related to process thermodynamics or reaction kinetics, although their accuracy remains somewhat obscure due to the approximate nature of the so-called exchange-correlation energy functional that must be selected.

In addition, its single-configuration nature, i.e., the fact that only one determinant represents the many-electron state, also prevents it from delivering accurate energies for systems that require a more complicated wave function ansatz as a superposition of many electronic configurations, i.e., many determinants beyond a single Slater determinant. This poses severe challenges for standard Kohn-Sham DFT in a wide range of strongly correlated systems such as molecular systems with one or multiple transition metals, bond breaking and transition states, light-matter interaction, and, in practice, may require the adoption of some sort of symmetry breaking (typically, that of spin symmetry)\cite{sinnecker2004calculating}.

The straightforward way of solving strongly correlated systems would be to expand the total wave function into a complete many-electron basis, i.e., as a linear combination of all possible Slater determinants that can be constructed in a given one-electron basis: 
\begin{equation}\label{eq:fci-wfn}
        \ket{\Psi} = c_0 \ket{\psi} + \sum_{ia} c_{ia} \ket{\psi_{i}^{a}} + \sum_{ijab} c_{ijab} \ket{\psi_{ij}^{ab}} + ... 
\end{equation}
with expansion coefficients $c$ (the so-called configuration interaction (CI) coefficients) that parametrize the state.
Inserting \cref{eq:fci-wfn} into \cref{eq:tise} turns the time-independent Schr\"odinger equation into a matrix eigenvalue problem,  
\begin{equation} \label{eq:eig}
    \textbf{HC} = \textbf{CE},
\end{equation}
where \textbf{H,C,E} are the matrix representations of the Hamiltonian, the CI coefficients, and energies, respectively. 
This is also referred to as ``Full Configuration Interaction'' (FCI).
In practice, the exact solution of the FCI problem is only possible for rather small chemical systems (i.e., those with less than about 18 spatial orbitals) on classical computers. \cite{fdez2019openmolcas} This is due to the exponential scaling of storing the wave function with respect to the number of orbitals, even when using subspace methods such as Lanczos or Davidson algorithms. \cite{lanczos1952,davidson1975}

\section{Quantum and traditional algorithms for molecular science}
\subsection{Energy evaluation} \label{sec:energy} 
\subsubsection{Quantum phase estimation and its traditional rivals}
As total electronic energies are the basis for any theoretical description of molecular systems, to calculate them with known accuracy is of decisive importance. Given that the exponentially scaling wall in FCI calculations can be overcome by quantum computing, we first discuss how quantum algorithms can deliver such exact energies (i.e., eigenvalues of the FCI problem in a given orbital basis). 

The quantum phase estimation (QPE) algorithm \cite{abrams1999quantum,nielson2000quantum} 
offers an alternative approach to solve the FCI problem on a quantum computer with a controllable error. Note the key feature that the error of the FCI energy (in a given one-electron basis) will be controllable for a specific system under consideration unlike in almost all traditional approaches to the electronic structure problem.
In QPE, one chooses a trial state $\ket{\Psi_{\mathrm{trial}}}$, a target error $\epsilon$ in the eigenvalue estimate, and a desired success probability $p$. 
The algorithm, which costs $n=\mathcal{O}(\frac{\text{poly}(N)}{\epsilon}\log(1/p))$ quantum gates, then returns an estimate $\hat{E}_j$ of a randomly selected eigenstate $H\ket{\Psi_j}=E_j\ket{\Psi_j}$.
This estimates satisfies
\begin{align}
    \textbf{Pr}\left[|\hat{E}_j-E_j|\le \epsilon\right]\ge 1-p.
\end{align}
Importantly, the eigenvalue $E_j$ 
is sampled with a probability $p_j=|\langle{\Psi_j}|\Psi_{\mathrm{trial}}\rangle|^2$. 
If ground state energies are desired, then $\ket{\Psi_{\mathrm{trial}}}$ should be chosen to make $p_0$ reasonably large.
%
%Additionally, the algorithm projects $\ket{\Psi_{\mathrm{trial}}}$ mostly onto the subspace of quantum states with energies $|\hat{E}_j-E_j|\le \epsilon$, with at most a fraction $p$ outside this subspace.
%
%This means that if the energy gap $\delta$ between this subspace and all other energy levels is large,

The performance of QPE may be compared to the classical power (subspace) iteration algorithm.
Power iteration multiplies $\ket{\Psi_{\mathrm{trial}}}$ by the Hamiltonian (shifted by the identity to have only negative eigenvalues)  $n$ times using $n e^{\mathcal{O}(N)}$ classical operations.
The resulting normalized quantum state is then $\ket{\Psi}\propto H^n\ket{\Psi_{\mathrm{trial}}}$. 
Given a target gap parameter $\delta$, and choosing $n=\mathcal{O}(\frac{1}{\delta}\log(1/p))$,
this state is guaranteed to have an overlap of at least $1-p$ with the subspace spanned by eigenstates of $H$ within $\delta$ of the ground state. If the energy gap of the Hamiltonian is larger than $\delta$, then exponential convergence of $p$ allows the ground state energy to be computed with logarithmic cost in error. 

On quantum computers, the storage requirements (i.e., the number of qubits) for the wave function is polynomial in the number of orbitals.  
But as a trade-off, the probability $p_j$ of sampling the desired quantum state $\ket{\Psi_j}$ means that QPE algorithms need to be repeated $\mathcal{O}(1/p_j)$ times, and the cost of the algorithm $n$ scales inversely with the precision of the energy estimate. 
In contrast, power iteration simply increases $n=\mathcal{O}(\frac{1}{\delta}\log(1/(p_jp)))$ to ensure large overlap with the desired subspace.
Moreover, once $\ket{\Psi}$ is computed, its expected energy $\bra{\Psi}H\ket{\Psi}$ may be computed exactly in a single step. In other words, QPE achieves an exponential improvement in $N$ for the number of operations required to obtain the energy of a randomly sampled eigenstate. 
The downside, however, is worse scaling with $p_j$ when a specific eigenstate is targeted.

Note that, although subspace FCI provides exact solutions for a quantum many-body problem in a given one-particle basis (be it the electronic or the nuclear Schr\"odinger equation in a basis of orbitals or modals, respectively) and it has
certain similarities to QPE, such a method should not be used as a benchmark or metric to assess the advantage of quantum algorithms. Subspace FCI is commonly not used in routine chemistry applications due to its very restricted size of the affordable one-particle basis and because of the fact that reliable relative energies may not require ultimate accuracy of total energies. In the spirit of FCI, many lower-scaling algorithms have been developed for chemistry application, for example
\begin{enumerate}
    \item Coupled-Cluster Singles and Doubles with Perturbative Triples (CCSD(T)) \cite{raghavachari1989,bartlett2007coupled}
    \item Complete-Active-Space Self-consistent Field (CAS-SCF) \cite{roos1980,ruedenberg1982atoms,gonzalez2020quantum}
    \item Density Matrix Renormalization Group Configuration Interaction/Self-consistent Field (DMRG-CI/SCF) \cite{white1992,white1999ab,baiardi2020density}
    \item Full Configuration Interaction Quantum Monte Carlo (FCIQMC) \cite{booth2009}
    \item Multi-Reference Configuration Interaction (MR-CI+Q) \cite{buenker1974,buenker1975,langhoff1974configuration,szalay2012}
    \item Multi-Reference Second Order Perturbation (MR-PT2) \cite{andersson1990,angeli2001,kurashige2011}
\end{enumerate}

These traditional algorithms represent the state of art
for solving the electronic Schr\"odinger equation to high accuracy, and we refer to refs.\citenum{motta2017towards,williams2020direct,eriksen2020ground} for detailed comparisons of these methods. Their core idea is still to solve the eigenvalue problem either using predefined restrictions of the many-electron basis (CCSD(T), MRCI+Q) or through an iterative construction of the basis-set expansion (DMRG, FCIQMC). 
As encoding the exact determinant space for many orbitals ($>$ 18) is hardly possible on classical hardware, a key aspect of all these novel methods is to approximate the full determinant space. 

The restriction to a selected finite set of so-called active orbitals in all FCI-type approaches generates a somewhat artificial distinction of electronic correlations into those that are called static (typically characterized by orbitals that occur in determinants with large weight in the wave function expansion) and those that are called dynamical (referring to orbitals present in determinants with small to vanishing weights). We note in passing that this artificial split into static and dynamical electron correlations can be overcome if a routine numerical approach becomes available to obtain results of FCI quality in a one-particle basis of one to a few thousand orbitals. Only quantum computing holds the promise to accomplish this goal, provided that a sufficiently large quantum computer can be built.

In CCSD(T), only the Hartree-Fock determinant and descendent determinants derived from single and double orbital-substitution operations are considered in the many-electron basis space (triple substitutions are added in a perturbative way). Such an approach allows for the treatment of a very large orbital space. Hence, both static and dynamical correlations can be recovered for a wide range of chemical systems.
CCSD(T) is considered the gold standard in traditional quantum chemistry methods. However, due to the single reference nature of the method, it still breaks down for systems with strong static correlations,
e.g. open-shell metal complexes, or non-equilibrium structures involving double-bond (and beyond) forming and breaking. For those challenging systems, where the static correlation is the key, one needs to use active space methods.

Unlike CCSD(T), the other aforementioned methods treat the chemical systems as potentially dominated by many determinants (i.e., in a multi-configurational fashion) and express this feature in terms of the choice of an active orbital space. CAS-CI is a pristine treatment of full configuration interaction in a subset of chemically most relevant orbitals (active space). DMRG-CI uses matrix product states to succinctly express the wave function. FCIQMC instead uses a coarse grained so-called walker distribution to sample the determinant space. For such active space methods, the focus is on recovering the static correlations originating from the multi-configurational nature of an electronic state under consideration. A follow-up orbital optimization can be added in so-called self-consistent field variants of these methods (i.e, CAS-SCF\cite{roos1980}, DMRG-SCF\cite{ghosh2008orbital}, and FCIQMC-SCF\cite{li2016}) to further minimize the total energy in a variational sense by finding a better one-particle basis for the restricted active space, i.e., CAS-CI expansion. 
Two approaches are typically applied to recover the missing dynamical correlations that arise from neglecting the major part of the virtual orbitals from the active space. One choice is to apply perturbation theory (usually to the second order, PT2) on top of the multi-configurational wave functions. This leads to methods such as CASPT2\cite{andersson1990}, NEVPT2,\cite{angeli2001} DMRG-PT2,\cite{kurashige2011} etc. 
The other approach is to apply a truncated-order configuration interaction (e.g. CISD) over the complete orbital space starting from a multi-configurational wave function (including corrections for size consistency), which leads to methods such as MRCI+Q. \cite{szalay2012}

DMRG and FCIQMC have become routine traditional FCI-type approaches in recent years as they can handle a much larger active space than the conventional CAS-CI algorithm at reasonable computing times. However, formally they still have exponentially scaling resource requirements for storing the wave function, just with a much smaller pre-factor. 
For instance, initiator-FCIQMC, which is the state-of-the-art FCIQMC method, holds a roughly $10^{-10}$ pre-factor in its exponential form of wave function storage for strongly correlated systems, which easily enables the approach to handle up to around 50 orbitals. 
In a system with relatively weak static correlation like benzene, a pre-factor of $10^{-25}$ has been achieved and led to a record active space size of 108 orbitals. \cite{ghanem2019}

In the near future, traditional methods such as DMRG and FCIQMC will remain more practical than quantum algorithms. However, we note that it may be difficult to rigorously assess the error in the energy after a fixed number of optimization cycles with pre-defined parameters (such as the bond dimension for DMRG or the number of walkers for FCIQMC).
Whether quantum computing will become competitive in the realm of FCI-type approaches will depend on advances in physical memory size and its communication speed with the CPU, 
because the polynomial scaling of storing the wave function will ultimately become the most significant advantage of quantum computing for the solution of problems in the molecular sciences.

\subsubsection{Variational Quantum Eigensolver}
Unfortunately, it is unfeasible to implement QPE on near-term quantum hardware due to its long runtime and the resulting need for large-scale, fault-tolerant quantum computing (requiring a huge number of physical qubits). An alternative approach, the so-called variational quantum eigensolver (VQE) \cite{mcclean2016theory}, is more suitable for near-term quantum hardware: Instead of running a single long calculation on a quantum computer in the case of QPE, VQE iteratively executes and optimizes a short parametrized quantum circuit that encodes the wave function ansatz. Due to the sampling involved in evaluating the energy of the ansatz at every optimization step, the scaling of VQE with the desired accuracy $\epsilon$ is  $\mathcal{O}(1/\epsilon^2)$, in contrast to $\mathcal{O}(1/\epsilon)$ for QPE.

In VQE, the parametrized wave function ansatz $\Psi(\theta)$ results in an expression for the energy of the form
\begin{equation} \label{eq:vqe}
    E(\theta) = \frac{\bra{\Psi(\theta)}H\ket{\Psi(\theta)}}{\braket{\Psi(\theta)}{\Psi(\theta)}} = \bra{\Psi(\theta)}H\ket{\Psi(\theta)}.
\end{equation}
Note that on a quantum computer the wave function is necessarily normalized so that we can ignore the normalization in the denominator of \cref{eq:vqe}. This expectation value is always larger than the smallest eigenvalue $E_0$ of $H$ owing to the variational principle. This allows one to use classical computers to optimize $\theta$ in order to find an approximation to $E_0$.

Intuitively, VQE can be understood as a direct analog of variational Monte Carlo (VMC)\cite{foulkes2001quantum}, with the differences that 1) the wave function $\Psi(\theta)$ is now stored on a quantum computer, so it retains the merit of polynomial scaling in storage, and 2) the energy is now evaluated through measurements instead of Monte Carlo integration. The variational optimization of wave function parameters $\theta$ is then done on classical computers just like in VMC. 

Its hybrid nature allows VQE to be implemented on a quantum device with much less coherence time. Unlike QPE, which gives the exact FCI energy, VQE gives a variational upper bound on the energy. The accuracy of VQE is thus limited by the ansatz it adopts. Multiple wave function ans\"atze have been investigated, developed, and implemented for VQE, e.g., Hartree-Fock (HF)\cite{google2020hartree}, unitary-coupled-cluster (UCC)\cite{evangelista2019exact}, qubit-coupled-cluster (QCC)\cite{ryabinkin2018qubit}, etc.
A good ansatz needs to be able to closely represent the exact ground state, require as few as possible iterations to find the parameter $\theta$ which minimizes $E(\theta)$, and have an efficient implementation on hardware. For a recent review on the different ans\"atze used in VQE, see ref.\citenum{bharti2021noisy}.

Once the energy of the ansatz has been measured, one can optimize the variational parameters on classical computers to look for the minimum. 
The hardness of the optimization problem depends on the chosen ansatz and the initialization. For example, \citet{mcclean2018barren} showed that there are cases with so-called ``barren plateaus'', i.e., the probability that the gradient along any reasonable direction is non-zero to some fixed precision is exponentially small as a function of the number of qubits. In ref.\citenum{bittel2021training}, Bittel and Kliesch constructed a free fermion problem, which is solvable in polynomial time, yet optimizing the variational parameters in a VQE formulation is NP-hard. 
%\todo{MT: Guang Hao and Hongbin, can you please add two more topics to the discussions: refer to my paper with Hastings and Wecker about the large number of measurements and refer to the paper from a couple years ago where a German group (including the BASF team) showed that more than $10^5$ gates are needed for chemcial accuracy at scale with current ansatz wave functions and that this is thus at an interesting scale beyond NISQ. VQE is thus nice for demos but at interesting scale QPE is preferred}

Though the energy evaluation of a couple of molecules have been demonstrated on NISQ hardware by VQE algorithms,\cite{o2019quantum,o2019calculating,nam2020ground,kandala2017hardware,google2020hartree,kawashima2021efficient} it must be noted that VQE does not scale well with respect to the molecule size. Recent work showed that to reach chemical accuracy with a UCCSD wave function ansatz at around 100 spin-orbitals, VQE requires $10^5$ gates\cite{kuhn2019accuracy}, which is already beyond what is feasible on NISQ hardware. Furthermore, the $10^{11}$ measurements per optimization step \cite{wecker2015progress} further increase the cost. Therefore, VQE is a solution for demonstration purposes on NISQ hardware, but QPE is to be preferred for routine real-world applications. 

\subsubsection{Exited state energy}
Another important direction is the evaluation of excited state energies. This is difficult for traditional quantum chemistry algorithms. Popular approaches like linear-response time-dependent DFT (TD-DFT) \cite{runge1984density} or algebraic diagrammatic construction (ADC)\cite{schirmer1982beyond} have introduced further approximations thus lowering their accuracy.\cite{laurent2013td,suellen2019cross} Even for the state-of-the-art equation-of-motion CCSD (EOM-CCSD)\cite{stanton1993equation} or iterative approximate coupled cluster singles, doubles, and triples  (CC3) \cite{christiansen1995response} methods, there still exist certain limitations. 
%CMR: I do not think that the sentence below is actually true and would delete it here:
%In particular, they can only probe singly excited states. \cite{schreiber2008benchmarks} 
Classically, CAS-CI, DMRG-CI, and MRCI+Q are among the very few methods that can resolve various excitation characters (singly, doubly, ..., n-tuply) in both low-energy (valence excitation) or high-energy ranges (core excitation). On the other hand, quantum algorithms are much more versatile when evaluating excited state energies. QPE can probe excited state energies in the same way as ground state energies.  The only requirement is to have an excited-state-like trial state, which can be prepared using MRCI or other wave function ans\"atze. \cite{bauman2020}  Though VQE was originally designed to solve the ground state only, there has been significant progress in applying VQE for excited states \cite{mcclean2017hybrid,higgott2019variational}. However, in contrast to QPE, VQE incurs additional overheads in resources and measurements in the case of excited states.

%\todo{wouldn't this be a better fit in the VQE subsection? or in a separate subsection after "excited state energy"?}
\subsubsection{Ansatz fidelity}
Both QPE and VQE need to adapt certain wave function ans\"atze to prepare the trial wave function. But the two methods have very different requirement in terms of the ansatz fidelity. QPE can reach unlimited precision using any initial state of non-negligible overlap with the true ground state. Previous research shows that one can use a $50\%$ fidelity ansatz to still obtain the FCI energy within chemical accuracy using QPE.\cite{bauman2020} In contrast, the precision of VQE depends directly on the trial state. To obtain chemical accuracy, it is necessary, through the choice of ansatz and tuning of variational parameters, to obtain 99.9\% fidelity or more with the true ground state wave function. This requirement itself already poses a challenge in the knowledge of efficient wave function ans\"atze, especially for strongly correlated systems.  Thus, we expect the development of VQE to also advance classical algorithms.

We summarize the key differences between QPE and VQE for energy evaluation of molecular systems in \cref{tab:qpe-vqe}. In terms of the resource requirement and cost of computing, we currently expect QPE to deliver the real quantum advantage for molecular systems in the long term. 

\begin{table}[H]
\footnotesize
\begin{tabular}{|c|c|c|}
\hline
                                 & QPE                                                                                    & VQE                                                                                                                                                        \\ \hline
Solution Accuracy                & \begin{tabular}[c]{@{}c@{}}Exact eigenstate\\ with controllable error\end{tabular}     & \begin{tabular}[c]{@{}c@{}}Limited by ansatz $\ket{\Psi(\theta)}$ \\ with uncontrollable error\end{tabular}                                                                   \\ \hline
Cost                             & $O(\#terms/\epsilon)$                                                                                   & $O((\#terms/\epsilon)^2)\times \text{optimization steps}$                                                                                                                                                       \\ \hline
Ansatz fidelity requirement & non-negligible & 99.9\% \\ \hline
Small non-trivial demonstrations & H$_2$ on NISQ                                                                             & H$_{10}$, H$_2$O on NISQ                                                                                                                                           \\ \hline
Industrial scale demonstrations  & \begin{tabular}[c]{@{}c@{}}Requires logical qubits\\ $10^{11}$ Toffoli gate operations\cite{vonberg2020}\end{tabular} & \begin{tabular}[c]{@{}c@{}}Requires logical qubits\\ $10^5$ gates for state prep\cite{kuhn2019accuracy}\\ $10^{11}$ measurement for energy evaluation\cite{wecker2015progress}\\ $10^3$ gradient descent\end{tabular} \\ \hline
\end{tabular}
\caption{Comparison of QPE and VQE algorithms}. %\todo{Please add references to the various claims for numbers for QPE and VQE}
\label{tab:qpe-vqe}
\end{table}

%%Property evaluation
\subsection{Chemical properties} \label{sec:property}
We have discussed how quantum algorithms can help obtain accurate electronic energies (for both, ground and excited states) of chemical systems, but many chemistry applications need properties beyond electronic energies. 

Free energy is one of the key quantities in thermodynamics and serves as the decisive measure for predicting reactions or drug-protein docking. Free energy may be approximated as a combination of separated degrees of freedom, i.e., as a sum of electronic, vibrational, rotational, and translational free energies. 

In traditional computational chemistry, one can use Monte Carlo sampling\cite{torrie1977nonphysical} or classical molecular dynamics \cite{sprik1998free,bussi2020using} 
to study an ensemble of molecular structures. This approach allows one to obtain all four free energy contributions and their couplings in one shot. However, due to the number of steps needed in both methods, one needs a rapid way to evaluate the energy. Force Field (FF) methods are therefore most frequently used in this approach. By virtue of low-scaling DFT and hybrid quantum-mechanics and molecular-mechanics (QM/MM) \cite{senn2009qm} methods, first-principles molecular dynamics has also been adopted in this field. \cite{lu2016qm}
Although these low-scaling or even empirical methods cannot provide the total energy to chemical accuracy, the free energy difference is a relative quantity where many errors may be expected to cancel. FF and DFT methods work reliably in  applications such as binding free energy predictions for drug-protein docking, although they are generally prohibitively expensive for high-throughput screening studies. 
On the quantum algorithm side, the quantum-Metropolis-Hastings algorithm can provide a quadratic speedup for Monte Carlo simulations.\cite{szegedy2004quantum,lemieux2020} However, the algorithm needs separated bases for electronic and vibrational degrees of freedom. While the electronic wave function can be prepared by various ans\"atze, it is much more challenging to prepare the vibrational basis for chemical systems\cite{bowman1978self, bowman1979application, christiansen2004vibrational,barone2005anharmonic,baiardi2017vibrational}, which makes the quantum acceleration of free energy calculations less straightforward\cite{sawaya2020resource,ollitrault2020hardware}.

Another routine application of quantum chemistry is to optimize and predict geometries of molecules. For most of the small and intermediate size molecules, chemical intuition can usually give good hints toward the approximate geometry. The geometry optimization in this case is usually to optimize to the nearest local minimum. This is a fairly simple task classically that can be done through various numerical methods such as quasi-Newton-Raphson or gradient descent. \cite{schlegel2011geometry} Additionally, the analytical gradient is generally available and reliable for FF and DFT methods; hence, a quantum algorithm like quantum gradient estimation \cite{jordan2005fast} does not provide any obvious advantage. For large soft molecules like proteins, where the global minimum geometry is of the most chemical interest, the geometry optimization becomes a non-trivial task. Grover search \cite{grover1997quantum} can provide a quadratic speedup compared to classical brute force search on the potential energy surface (PES). However, it is not a practical approach as classically the geometry optimization will not be conducted in a complete black box way. Heuristics like genetic algorithms\cite{deaven1995molecular} or simulated annealing \cite{mundim1996geometry} are often used to search for global minimum geometry. In this case, quantum walks \cite{szegedy2004quantum,lemieux2020} may replace classical random walks in simulated annealing to provide a quadratic speed up. However, due to the faster clock speed of classical computers, quadratic speedups are unlikely to be practical in the foreseeable future. %\todo{MT: I suggest to emphasize the exponential versus quadratic issue once more in the conclusions}.

%%Dynamics
\subsection{Dynamics} \label{sec:dynamics}
There are certain chemical phenomena that can only be studied through explicit time evolution. For instance, in photo-active systems, one wants to study on which time scale internal conversion and intersystem-crossing occurs. Answering such dynamics questions requires direct solution of \cref{eq:tdse}. In traditional quantum chemistry, methods like surface hopping \cite{martinez1997molecular,ben2000ab,schmidt2008mixed,subotnik2016understanding}, 
and multi-configurational time dependent Hartree (MCTDH) \cite{meyer1990multi,beck2000multiconfiguration,manthe2008multilayer} 
have been frequently used to explicitly propagate nuclear and electronic wave packets over time. Autocorrelation functions then become key quantities for information extraction. One of the prerequisites for those traditional methods to work well is to have an accurate representation of the PES. This is essentially an electronic structure problem as discussed in \cref{sec:energy}. On quantum computers, such autocorrelation functions are no longer needed, because Hamiltonian time-evolution is simulated directly. 
This means that in order to solve a time-resolved problem such as determining when an intersystem-crossing happened, one will need to run multiple Hamiltonian simulations for different time periods to search for, e.g., a singlet/triplet population crossing point. Binary search can be used to reduce the cost. Another challenge of applying quantum algorithms for dynamics is that one needs to carefully think what to measure given the limited information one can obtain from a wave function. 
%\todo{MT: designing the quantum circuit for evolution should not depend on the measurements. I assume you mean that given the limited information one can obtain from a wave function one should think carefully about what to measure?}

%%Cheminformatics
\subsection{Cheminformatics} \label{sec:cheminformatics}
Over the years, the development of modern computer architecture and advancements in quantum chemistry software have made cheminformatics another booming application, especially for screening drug and material candidates. \cite{hartenfeller2010novo,gasteiger2016chemoinformatics}

Machine learning, as a data-driven approach, is expected to generate new insights from massive existing quantum chemistry data. Several machine learning applications have been developed to extract highly accurate predictions from low-precision quantum chemistry calculations using various neural network models. \cite{carleo2017solving,cheng2019universal,chen2020ground,dick2020machine,chen2020deepks,hermann2020deep}

Moreover, there is growing interest in applying quantum machine learning techniques to change the landscape of cheminformatics.\cite{biamonte2017quantum,von2018quantum,bharti2020machine,christensen2019operators} However, quantum machine learning is still in its infancy and  there are known challenges to quantum computers reading classical data\cite{preskill2018quantum}. It is unclear whether quantum machine learning models offer a practical advantage over classical approaches.  In the near future, we therefore still expect classical machine learning to be dominantly used in cheminformatics. %\todo{MT: a speculation that we could add is that QML could be useful in learning quantum wave functions.} 
However, one speculation is that quantum machine learning may help learn the compact form of wave function as in this case, only Hamiltonian parameters are needed for the learning. This kind of problem will read the same small amount of classical data just like solving a typical molecular energy problem. \cite{carleo2017solving}

\section{Conclusions and Perspectives}

In this work, we provided an overview of potential target application areas in the molecular sciences for quantum computing. We highlighted the competition with state-of-the-art traditional methods. Especially in light of the tremendous achievements of traditional algorithms on classical computers, demonstrating a game-changing quantum advantage is a complicated and multi-faceted task. Current achievements in actual quantum computation are impressive and truly encouraging, but we still have a long path ahead of us. 

A quantum algorithm may be shown to have a formal advantage in terms of scaling and efficiency over some traditional approaches. However, recently we have started to realize only super-quadratic speedup quantum algorithms have the potential to excel the classical algorithms as there is a big constant speed advantage of classical computers. Even for quantum algorithms with exponential speedups it is not clear whether they result in an advantage over the best available traditional algorithms for a given practical problem once all overheads are taken into account.
In order to define what kind of advancement has been achieved in the future, it could be helpful to introduce metrics that can measure its performance against the best state-of-the-art approaches and that relate to actual computations on a quantum machine. 

Potential criteria to keep in mind for such metrics may be the following: 
\begin{enumerate}
    \item Serial speed: For a given computational task, assess the time required on a state-of-the-art classical computer for the fastest traditional method that can deliver the same precision as the quantum algorithm on a quantum computer under the same constraints (e.g., same one-particle basis set), for which one either measures or estimates a time. One may divide the former measured/estimated time by the latter to obtain a speedup ratio. Note that the target precision may vary depending on the accuracy needed for the computation result, which depends on its scientific purpose. 
    \item Parallel speed: Accordingly, for a given scientific target, measure the reduction in computing time when parallelization is taken into account without changing any of the settings defined for the serial speed consideration above.
    \item Cost: One may consider the cost in terms of computer acquisition, life time, energy consumption, and environmental impact to decide whether a potentially more inefficient traditional calculation might be preferred. 
    \item Accessibility: One would like to know how easy it will be to access a quantum machine for an ordinary scientist (e.g., everybody can do traditional quantum calculations on a laptop these days) 
    \item Scalability: The size of the problem, given, e.g., by the size and type of the molecule which sets the number of orbitals to be considered, will require flexible hardware that can cope with this changing parameter. 
\end{enumerate}

In recent years, we have seen remarkable achievements in quantum hardware and algorithm development
and there is no reason to believe that this accelerating pace will be slowing down any time soon. However, the
metrics mentioned above will ultimately decide on the success and fate of the different strategies 
that are currently being pursued in the field of quantum computation for molecular science.

%\section*{Availability of data and material}
%Not applicable.

%\section*{Competing interest}
%The authors declare that they have no competing interests.

%\section*{Funding}
%Not applicable. 

%\section*{Authors' contributions}
%H.L., G.H.L, D.S.S., T.H., M.R., and M.T wrote the manuscript together.

\begin{acknowledgement}
The authors thank Stephen Jordan and Martin Roetteler for stimulating discussions.
\end{acknowledgement}

\bibliography{acs-achemso}

\end{document}